\documentclass[article,aps,draft,showpacs,showkeys]{revtex4}

\begin{document}
\preprint{APS/123-QED}

\title{A General Scheme for Construction of Coherent States of 
Anharmonic Oscillators}

\author{Marcin Molski}

\affiliation{
Department of Theoretical Chemistry, Faculty of Chemistry\\
A. Mickiewicz University of Pozna\'n,\\
ul. Grunwaldzka 6, PL 60-780 Pozna\'n, Poland}

\email{marcin@rovib.amu.edu.pl}

\date{\today}

\begin{abstract}

A mixed supersymmetric-algebraic approach to construction of the minimum uncertainty 
coherent states of anharmonic oscillators is presented. It permits generating not only the well-known 
coherent states of the harmonic and Morse oscillators but also the so far unknown coherent states 
of the Wei Hua, Kratzer-Fues and generalized Morse and Kratzer-Fues oscillators. The method 
can be applied also to generate superpotentials indispensable for deriving the Schr\"odinger 
equation in the  supersymmetric form amenable to direct solution in the SUSYQM scheme. 

\end{abstract}

\pacs{03.65.Fd}

\keywords{Algebraic methods,  Anharmonic oscillators, Coherent states, Quantum supersymmetry}

\maketitle

The coherent states introduced by Schr\"odinger in 1926 \cite{S}, make a very useful tool for investigation of 
various problems in quantum optics \cite{Glauber}, in particular the interactions of matter with coherent 
radiation  \cite{Walker}. Such resonant interactions produce the coherent effects, e.g. self-induced 
transparency, soliton formation, excitation of a coherent superposition of rotational states, periodic 
alternations of the refractive index in both nuclear and molecular systems \cite{Walker}. 
Studies in the latter case require construction of the coherent states for 
anharmonic molecular potentials. Such states are defined in the similar manner as the ordinary
coherent states of the harmonic oscillator \cite{Zhang}: (i) they are eigenstates of the annihilation operator, 
(ii) they minimize the generalized position-momentum uncertainty relation and (iii) they arise from the operation 
of a unitary displacement operator to the ground state of the oscillator. It should be pointed out that the definition 
(iii) relies on the form of the displacement operator, which is specific to harmonic oscillator \cite{Cooper}, hence, 
in this case mainly approximate coherent states can be derived using, for example, Nieto and Simons 
\cite{Nieto} or Kais and Levine \cite{Kais} procedures.

Coherent states of anharmonic oscillators have been constructed using several alternative approaches. In the method 
proposed by Nieto and Simmons \cite{Nieto}, the position and momentum operators are chosen in such a way that
the resultant Hamiltonian resembles that for an harmonic oscillator. The coherent states are then
determined on condition that they minimize the generalized uncertainty relation in the new variables.  
Perelomov \cite{Perelomov} has derived the coherent states using the irreducible representations of 
a Lie group. This method has been successfully applied to generate the coherent states of the  
Morse oscillator \cite{Kais}. The generalized coherent states can also be generated using an 
algebraic method \cite{Cooper} and supersymmetric quantum mechanics (SUSYQM) \cite{Fukui}.
Applying the above formalisms, the coherent states for Morse \cite{Cooper, Nieto, Kais}, P\"oschl-Teller \cite{Teller}, 
hydrogen atom \cite{H}, Eckart and Rosen-Morse \cite{Fakhri}, double-well and linear (gravitational) potentials 
\cite{Nieto1} have been constructed. 

In the present study, we introduce the mixed supersymmetric-algebraic 
method, which permits generating not only the coherent states of the harmonic and Morse oscillators but also
the so far unknown coherent states of the Wei Hua \cite{Wei}, Kratzer-Fues \cite{Kratzer} and generalized Morse 
and Kratzer-Fues oscillators. The method starts from the vibrational dimensionless Schr\"odinger equation
\begin{equation}
\left[\frac12\hat{p}^2+V(q)-E_0\right]|v>=\Delta E_{v0}|v>,\hskip1cm \hat{p}=-i\frac{d}{dq}
\label{Eq0}
\end{equation}
in which $\Delta E_{v0}=E_v-E_0$ whereas $q=u_rr$ denotes a dimensionless spatial variable $r$, with a scalling 
factor $u_r$ depending on the explicit form of the potential energy term $V(q)$. 

The crucial for the approach proposed is the assumption that the last two terms in the operator part of Eq. (\ref{Eq0})
can be specified in the form of the Riccati equation
\begin{equation}
V(q)-E_0=\frac12\left[x^2(q)+\frac{dx(q)}{dq}\right]
\label{Eq00}
\end{equation}
familiar in SUSYQM \cite{FCooper}. Here, $x(q)$ is the anharmonic 
coordinate, which satisfies the commutation relation $\left[x(q), \hat{p}\right]=i{dx(q)}/{dq}$.
Its form depends on the oscillator type, hence the explicit expression for $x(q)$ will be determined 
for a given form of the potential energy function. In SUSYQM $x(q)$ (with accuracy to sign) is interpreted 
as a superpotential \cite{FCooper}, which permits construction of the supersymmetric Schr\"odinger equation
straightforward to analytical solutions. 

Substituting Eq. (\ref{Eq00}) into (\ref{Eq0}) one gets the latter in the factorized form 
\begin{equation}
\hat{A}^{\dag}\hat{A}|v>=\Delta E_{v0}|v>,
\label{Eq1}
\end{equation}
in which 
\begin{equation}
\hat{A}= \frac{1}{\sqrt{2}}\left[\frac{d}{d q}- x(q)\right],\hskip.5cm 
\hat{A}^{\dag}= \frac{1}{\sqrt{2}}\left[-\frac{d}{d q}-x(q)\right],\hskip.5cm 
\left[\hat{A}, \hat{A}^{\dag}\right]=-\frac{dx(q)}{dq}.
\label{Eq3}
\end{equation}
In order to construct the coherent state for the potential $V(q)$, we need a ground state 
solution $|0>$ of Eq. (\ref{Eq1}), which is an eigenstate of the operator $\hat{A}$. If  $\hat{A}$
annihilates the ground state $\hat{A}|0>=0$, then the coherent states $|\alpha>$ are the eigenstates 
of the annihilation operator $\hat{A}$ and the following relations are fulfilled
\begin{equation}
\hat{A}|\alpha>=\alpha|\alpha>,\hskip.5cm<\alpha|\alpha^*=<\alpha|\hat{A}^{\dag},
\hskip.5cm |\alpha>=|0>\exp(\sqrt{2}\alpha q).
\label{Eq6}
\end{equation}
One may prove that the states $|\alpha>$ minimize the generalized position-momentum 
uncertainty relation \cite{Cooper}
\begin{equation}
(\Delta x(q))^2(\Delta p)^2 \ge\frac14{<\alpha|g(q)|\alpha>^2}\hskip.5cm 
g(x)=-i[x(q),\hat{p}]=\frac{dx(q)}{dq}=-\left[\hat{A}, \hat{A}^{\dag}\right].
\label{Eq7}
\end{equation}
To prove this thesis let's calculate
\begin{equation}
<\alpha|x(q)|\alpha>=\frac{1}{\sqrt{2}}<\alpha|\hat{A} + 
\hat{A}^{\dag}|\alpha>=
\frac{1}{\sqrt{2}}(\alpha+\alpha^*),
\label{Eq8}
\end{equation}
\begin{equation}
<\alpha|\hat{p}|\alpha>=-i\frac{1}{\sqrt{2}}<\alpha|\hat{A} - 
\hat{A}^{\dag}|\alpha>=
-i\frac{1}{\sqrt{2}}(\alpha-\alpha^*),
\label{Eq9}
\end{equation}
\begin{equation}
2<\alpha|x(q)^2|\alpha>=<\alpha|\hat{A}\hat{A} 
+2\hat{A}^{\dag}\hat{A} +
\hat{A}^{\dag}\hat{A}^{\dag}-\frac{dx(q)}{dq}|\alpha>=\left[(\alpha+\alpha^*)^2
-<\alpha|\frac{dx(q)}{dq}|\alpha>\right],
\label{Eq10}
\end{equation}
\begin{equation}
-2<\alpha|\hat{p}^2|\alpha>=<\alpha|\hat{A}\hat{A} -
2\hat{A}^{\dag}\hat{A}+
\hat{A}^{\dag}\hat{A}^{\dag}+\frac{dx(q)}{dq}|\alpha>=\left[(\alpha-
\alpha^*)^2+<\alpha|\frac{dx(q)}{dq}|\alpha>\right],
\label{Eq11}
\end{equation}
in which Eq. (\ref{Eq3}) is employed in the form
$\hat{A}\hat{A}^{\dag} = \hat{A}^{\dag}\hat{A}-{dx(q)}/{dq}$.

Having calculated  (\ref{Eq8}) - (\ref{Eq11}), we can pass to evaluate 
\begin{equation}
(\Delta x(q))^2=<\alpha|x(q))^2|\alpha> - 
<\alpha|x(q)|\alpha>^2=-\frac12<\alpha|\frac{dx(q)}{dq}|\alpha>,
\label{Eq13}
\end{equation}
\begin{equation}
(\Delta p)^2=<\alpha|\hat{p}^2|\alpha> - 
<\alpha|\hat{p}|\alpha>^2=-\frac12<\alpha|\frac{dx(q)}{dq}|\alpha>
\label{Eq14}
\end{equation}
providing $\Delta x(q)=\Delta p$ and 
\begin{equation}
(\Delta x(q))^2(\Delta p)^2=\frac14<\alpha|\frac{dx(q)}{dq}|\alpha>^2.
\label{Eq15}
\end{equation} 
The calculations performed prove that the states $|\alpha>$ minimize the generalized position-momentum 
uncertainty relation for the anharmonic coordinate $x(q)$. They are also the eigenstates of the operator 
$\hat{A}$, which annihilates the ground state $\hat{A}|0>=0$, hence they satisfy the two fundamental requirements 
established for the coherent states of an anharmonic oscillator.

In order to demonstrate how the method works, let's calculate first the well-known coherent states 
of the harmonic oscillator. To this purpose let's assume that       
\begin{equation}
\frac{dx(q)}{dq}=-1 \Longrightarrow x(q)=-q\hskip.3 cm{\rm for}\hskip.3cm x(0)=0. 
\label{Eq16}
\end{equation} 
Then Eq.(\ref{Eq0}) including (\ref{Eq00}) turns out to be the well-known Schr\"odinger equation for 
the ground state $|0>=\exp(-q^2/2)$ of an harmonic oscillator, whereas the operators 
$\hat{A}$ and $\hat{A}^{\dag}$
take the well-known form of annihilation and creation operators, which satisfy the 
commutation rule $ \left[\hat{A}, \hat{A}^{\dag}\right]=-{dx(q)}/{dq}=1$. Hence, 
the coherent states of an harmonic oscillator can be specified by 
the general formula (\ref{Eq6})
\begin{equation}
|\alpha>=|0>\exp(\sqrt{2}\alpha q)=\exp(-\frac12q^2)\exp(\sqrt{2}\alpha q).
\label{Eq18}
\end{equation}
The results obtained indicate that crucial for the method proposed is the 
explicit form of the term $dx(q)/dq$, which in the general case can be given 
as a negative $x$-dependent function
\begin{equation}
\frac{dx(q)}{d q}=-f(x).
\label{Eq21}
\end{equation}
Hence, employing different analytical functions $f(x)$ one may generate a variety of potentials and 
associated superpotentials satisfying relation (\ref{Eq00}). The indispensable for this purpose 
$x(q)$ can be calculated from (\ref{Eq21}) by integration, provided that we known the
explicit form of $f(x)$. Assuming that $|x(q)|<1$ one may expand $f(x)$ in a power series 
\begin{equation}
f(x)=c_1(x+c_0/c_1)+c_2(x+c_0/c_1)^2+.......
\label{Eq22}
\end{equation}    
and then successively apply the first-, second- and higher-order terms in determination 
of coherent states of anharmonic potentials. For example, if we take into account the linear 
term $f(x)=c_1(x+c_0/c_1)$ and assuming $x(0)=(1-c_0)/c_1$ one may calculate from 
(\ref{Eq21})
\begin{equation}
x(q)=\frac{1}{c_1}\left[\exp(-c_1q)-c_0\right]. 
\label{Eq24}
\end{equation}
Introducing (\ref{Eq24})  into (\ref{Eq00}) and redefining the constants $c_0=s-x_e$, $c_1=\sqrt{2x_e}$ from  Eqs. (\ref{Eq0}) 
and (\ref{Eq00})  one gets the Schr\"{o}dinger equation 
\begin{equation}
\frac12\left\{-\frac{d^2}{dq^2}+\frac{1}{2x_e}\left[s-\exp(-\sqrt{2x_e}q)\right]^2-s+\frac{x_e}{2}\right\}|0>=0
\label{Eq25}
\end{equation}
and the ground state solution
\begin{equation}
|0>=\exp[-\frac{1}{2x_e}\exp(-\sqrt{2x_e}q)]\exp\left[-\frac{1}{\sqrt{2x_e}}(s-x_e) \right]
\label{Eq25a}
\end{equation}
of the generalized Morse oscillator \cite{Morse} $V(r)=D_e\left[s-\exp(-ar) \right]^2$.
For $s=1$ the above specified formulae are reduced to the well-known 
equations derived by Cooper \cite{Cooper}, in which
$x_e={\hbar\omega_e}/{4D_e}$ denotes the anharmonicity constant, 
$\omega_e=a\sqrt{{2D_e}/{m}}$ is the vibrational frequency defined by the reduced mass $m$ 
of the system, the dissociation energy $D_e$ and the range 
parameter $a$ appearing in the Morse potential.

Eqs. (\ref{Eq24}) and (\ref{Eq3}) produce the generalized Morse annihilation and creation operators 
\begin{equation}  
\hat{A}=\frac{1}{\sqrt{2}}\left[\frac{d}{d q}+ \frac{\left(s-e^{-\sqrt{2x_e}q}\right)}{\sqrt{2x_e}}-\sqrt{\frac{x_e}{2}}\right],
\hat{A}^{\dag}= \frac{1}{\sqrt{2}}\left[-\frac{d}{d q}+ \frac{\left(s-e^{-\sqrt{2x_e}q}\right)}{\sqrt{2x_e}}-\sqrt{\frac{x_e}{2}}\right]
\label{Eq26}
\end{equation}
and associated coherent states 
\begin{equation}
|\alpha>=\exp\left[-\frac{1}{2x_e}\exp(-\sqrt{2x_e}q)\right]
\exp\left[ -\frac{1}{\sqrt{2x_e}}\left( s-x_e \right)\right]\exp(\sqrt{2}\alpha q).
\label{Eq28}
\end{equation}
They are eigenstates of the annihilation operator $\hat{A}$, which minimize the uncertainty relation (\ref{Eq7})
for $\left[\hat{A}, \hat{A}^{\dag}\right]=\exp(-\sqrt{2x_e}q)$.

Taking into account the parabolic expansion 
\begin{equation}
f(x)=c_1(x+c_0/c_1)+c_2(x+c_0/c_1)^2
\label{Eq30}
\end{equation}
and the identical assumption as previously $x(0)=(1-c_0)/c_1$,  from Eq.(\ref{Eq21}) one obtains
\begin{equation}
x(q)=\frac{(cc_1/c_2)\exp[-c_1(q-q_0)]}{1-c\exp[-c_1(q-q_0)]}-\frac{c_0}{c_1} 
\label{Eq32}
\end{equation}
in which $c=C/(B/W-C)$, $q_0=\ln(B/W-C)/c_1$, $W=(2c_0+c_1^2)/[2c_1(1-c_2)]$, 
$B=c_1/(c_1^2+c_2)$ and $C=c_2/(c_1^2+c_2)$. Hence, the Schr\"odinger equation (\ref{Eq1})
and its ground state solution take the forms
\begin{equation}
\frac12\left\{-\frac{d^2}{dq^2}+2D\left\{\frac{1-\exp[-c_1(q-q_0)]}{1-c\exp[-c_1(q-q_0)]}\right\}^2-2E_0\right\}|0>=0,
\label{Eq33}
\end{equation}
\begin{equation}
|0>=\left\{1-c\exp[-c_1(q-q_0)]\right\}^{\frac{1}{c_2}}\left\{c\exp[-c_1(q-q_0)]\right\}^{\frac{c_0}{c_1^2}},
\label{Eq34}
\end{equation}
in which $2D=(1-c_2)W^2$, $2E_0=W^2-c_0^2/c_1^2$. Eq. (\ref{Eq33}) is a well-known quantal equation for
the ground state of the Wei Hua oscillator \cite{Wei}, whose coherent states have not been derived as yet. 
Applying the above specified procedure one gets the correspondig annihilation and creation operators 
\begin{equation}
\hat{A}=\frac{1}{\sqrt{2}}\left[\frac{d}{d q}-\frac{(cc_1/c_2)\exp[-c_1(q-q_0)]}{1-c\exp[-c_1(q-q_0)]}+\frac{c_0}{c_1} \right],
\label{Eq35}
\end{equation}
\begin{equation}
\hat{A}^{\dag}= \frac{1}{\sqrt{2}}\left[-\frac{d}{d q}-\frac{(cc_1/c_2)\exp[-c_1(q-q_0)]}{1-c\exp[-c_1(q-q_0)]}+\frac{c_0}{c_1} \right],
\label{Eq35a}
\end{equation}
as well as the coherent states of the Wei Hua oscillator 
\begin{equation}
|\alpha>=\left\{1-c\exp[-c_1(q-q_0)]\right\}^{\frac{1}{c_2}}\left\{c\exp[-c_1(q-q_0)]\right\}^{\frac{c_0}{c_1^2}}
\exp(\sqrt{2}\alpha q).
\label{Eq37}
\end{equation}
They are eigenstates of the annihilation operator $\hat{A}|\alpha>=\alpha|\alpha>$ and minimize the uncertainty relation (\ref{Eq7})
for $\left[\hat{A}, \hat{A}^{\dag}\right]=(cc_1^2/c_2)\exp[-c_1(q-q_0)]/\{ 1-c\exp[-c_1(q-q_0)]   \}^2$.

The method proposed also permits a derivation of the coherent 
states of the Kratzer-Fues potential \cite{Kratzer} specified in the form
\begin{equation}
V(r)=D_e\left[\frac{r-r_e}{r}\right]^2=D_e\left[\frac{z}{1+z}\right]^2
\label{Eq38}
\end{equation}
in which $r_e$ denotes the equilibrium configuration of the molecule bond $V(r_e)=0$, whereas 
$z=(r-r_e)/r_e$ is the Dunham variable.
To this aim we employ the generating function $f(x)=[c_1(x+c_0/c_1)]^2$, which introduced 
into Eq. (\ref{Eq21}) yields
\begin{equation}
x(q)=\frac{1}{c_1(c_1q+1)}-\frac{c_0}{c_1},\hskip1cm x(0)=(1-c_0)/c_1.
\label{Eq39}
\end{equation}
Then the Schr\"{o}dinger equation (\ref{Eq0}) and its ground state solution for $x(q)$ 
specified above take the form
\begin{equation}
\frac12\left\{-\frac{d^2}{dq^2}+2D\left[\frac{c_1q-s}{1+c_1q}\right]^2-2E_0\right\}|0>=0,\ 
|0>=(1+c_1q)^{\frac{1}{c_1^2}}\exp\left[-\frac{(1-c_1^2)(c_1q+1)}{c^2_1(s+1)}\right],
\label{Eq41}
\end{equation}
in which $2D=c_0^2/[c_1^2(1-c_1^2)]$, $s=(1-c_0-c_1^2)/c_0$, $2E_0=c_0^2/(1-c_1^2)$. 
It is easy to verify that for $s=0$ or $c_0=1-c_1^2$ and $c_1q=z$,  Eq. (\ref{Eq41}) turns out to be 
the Schr\"{o}dinger equation for the Kratzer-Fues oscillator \cite{Kratzer}
\begin{equation}
\frac12\left\{\frac{d^2}{dq^2}+2D\left[\frac{c_1q}{1+c_1q}\right]^2-2E_0\right\}|0>=0,
\hskip.5cm
|0>=(1+c_1q)^{\frac{1}{c_1^2}}\exp[-(1-c_1^2)(c_1q+1)/c^2_1],
\label{Eq43}
\end{equation}
in which $2D=(1-c_1^2)/c_1^2$, $2E_0=1-c_1^2$. 
Hence the Kratzer-Fues annihilation and creation operators can be given in the form
\begin{equation}  
\hat{A}=\frac{1}{\sqrt{2}}\left[\frac{d}{d q}-\frac{1}{c_1(c_1q+1)}+\frac{1-c_1^2}{c_1}\right],
\hat{A}^{\dag}= \frac{1}{\sqrt{2}}\left[-\frac{d}{d q}-\frac{1}{c_1(c_1q+1)}+\frac{1-c_1^2}{c_1}\right]
\label{Eq44}
\end{equation}
and the associated coherent states 
\begin{equation}
|\alpha>=(1+c_1q)^{\frac{1}{c_1^2}}\exp[-(1-c_1^2)(c_1q+1)/c^2_1]\exp(\sqrt{2}\alpha q)
\label{Eq46}
\end{equation}
minimize the uncertainty relation (\ref{Eq7}) for $\left[\hat{A}, \hat{A}^{\dag}\right]={(c_1q+1)^{-2}}$.

The potential $V(z)=D_e[(z-s)/(1+z)]^2$, which appears in (\ref{Eq41}) is worth considering as it represents a
generalized version of the Kratzer-Fues formula $V(r)=D_e[1-r_e(s+1)/r]^2$. From the relation
$-1\le 1-r_e(s+1)/r\le 1$ one may calculate the convergence radius ${R}\in [r_e(s+1)/2, \infty]$ for the new
potential. The former increases for $s\in(-1,0)$ in comparison to the original Kratzer-Fues potential ($s=0)$ yielding
${R}\in (r_e/2,\infty)$. In such circumstances, the expansion of the potential energy function 
\begin{equation}
V(r)=c_0\left[\frac{r-r_e(s+1)}{r}\right]^2\left\{1+\sum_{n=1}^Nc_n\left[\frac{r-r_e(s+1)}{r}\right]^n\right\}
\label{Eq47}
\end{equation}
into a series of the generalized Kratzer-Fues variable, will provide much accurate reproduction of the real 
potential curves then that obtained by the Simons-Parr-Finlan expansion ($s=0$ ) \cite{SPF}, which diverges 
in the united-atom limit $r\to 0$. The set of parameters $(r_e,s,c_0,c_1,...)$ can be evaluated from the molecular IR 
and MW spectra by making use of the fitting procedure. It should be pointed out also 
that the new potential permits analytical solution of the Schr\"odinger equation and can be 
used to generate the coherent states for the generalized Kratzer-Fues oscillator 
\begin{equation}
|\alpha>=(1+c_1q)^{\frac{1}{c_1^2}}\exp[-(1-c_1^2)(c_1q+1)/c^2_1(s+1)]\exp(\sqrt{2}\alpha q).
\label{Eq48}
\end{equation}
The method proposed is general and permits construction of the coherent states, associated 
potentials and superpotentials as well as deriving the supersymmetric Schr\"odinger equation 
amenable to direct solution in the SUSYQM scheme \cite{Schwabl}.  In the standard approach the
superpotentials are solutions of the Riccati equation (\ref{Eq00}) obtained for the specific form 
of the potential function $V(q)$ \cite{Schwabl}. Here a new procedure has been introduced, 
which permits simultaneous derivation of the potentials and associated superpotentials 
assuming that ${dx(q)}/{dq}=-f(x)$. This term can be expanded in a power series of $x(q)$ 
and then used to generate the coherent states for different orders of the expansion (\ref{Eq22}). 
For the terms up to the second order, the method produces the minimum uncertainty coherent 
states for harmonic, Morse, Wei Hua, Kratzer-Fues and generalized Morse and Kratzer-Fues 
potentials. They are the most important potential energy functions employed in molecular 
quantum theory, theoretical spectroscopy and quantum optics.

\end{document}